\documentclass[tightenlines,superscriptaddress,eqsecnum,floats,nofootinbib,showpacs]
{revtex4-1}

\usepackage{amssymb}
\usepackage{stmaryrd}
\usepackage{amsmath}
\usepackage{amsfonts}
\usepackage{mathrsfs}
\usepackage{CJK}
\usepackage{amsmath,amssymb,amsfonts}
\usepackage{graphicx}

\def\be{\begin{equation}}
\def\ee{\end{equation}}
\def\ba{\begin{eqnarray}}
\def\ea{\end{eqnarray}}
\def\nn{\nonumber}

\newcommand{\abs}[1]{{\left|{#1}\right|}} 

\begin{document}


\title{ One possible solution of peculiar type Ia supernovae explosions caused by a charged white dwarf}

\author{Helei Liu}
\affiliation{School of Sciences, South China University of
Technology, Guangzhou 510641, P.R. China}

\author{ Xiangdong Zhang\footnote{Corresponding author. scxdzhang@scut.edu.cn}}
\affiliation{School of Sciences, South China University of
Technology, Guangzhou 510641, P.R. China}

\author {Dehua Wen }
\affiliation{School of Sciences, South China University of
Technology, Guangzhou 510641, P.R. China}

\date{\today}


\begin{abstract}
Recent astrophysics observation reveals the existence of some
super luminous type Ia supernovae. One natural explanation of such
a peculiar phenomenon is to require the progenitor of such a supernova
to be a highly super-Chandrasekhar mass white dwarf. Along this
line, in this paper, we propose a possible mechanism to explain
this phenomenon based on a charged white dwarf. In particular, by
choosing suitable new variables and a representative charge
distribution, an analytic solution is obtained. The stability
issue is also discussed, remarkably, it turns out that the charged white dwarf configuration can be dynamically stable.
Moreover, we investigate the general relativistic effects and it is  shown that the general relativistic effects can be negligible when
the mass of the charged white dwarf is below about $3M_\odot$.
\end{abstract}


\maketitle

\section{Introduction}

It is generally believed that type Ia supernovae can be regarded as standard candles in the measurement of cosmological distance.
 However, some peculiar type Ia supernovae, e.g., SN2003fg, SN2006gz, SN2007if and
SN2009dc \cite{Howell06,Scalzo10} have been observed  recently.
Contrast with the standard Ia supernovae, these objects have
exceptionally higher luminosity but with lower kinetic energies.
In order to understand these phenomenon, some authors assume that
the progenitor of such supernova to be a highly
super-Chandrasekhar mass white dwarf. If the masses of  progenitors
of above supernovae lie in the range $2.1M_\odot\sim 2.8M_\odot$
\cite{Howell06,Scalzo10,Hicken07,Yamanaka09,Silverman11,Taubenberger11},
here $M_\odot$ being the mass of sun, this hypothesis can explain
experiment data naturally. Hence along this line, the authors of
references \cite{Mukhopadhyay12,Das12,Das13} propose a mechanism to
generate super-Chandrasekhar mass white dwarf by assuming the
existence of a super strong uniform magnetic fields inside the
white dwarf. However, it becomes a debate issue, although this picture can uplift the mass of
white dwarf significantly, it questioned by several authors with its stability
issues \cite{Zuo14}, including neutronization induced by inverse
beta decay \cite{CFD13}, dynamical instability as demonstrated
by literature \cite{Ruffini13}. Reply of those criticism can be found in\cite{DM14a,DM14b}. Now a
natural question is that does it exist any other possibility to
form a highly super-Chandrasekhar white dwarf? Can we find a
stable configuration significant violating the Chandrasekhar mass
limit? If it exists, then what is the ultimate mass limit of a
white dwarf? Here we plan to address all the above issues by
exploiting the electric field effects inside the compact star.

In fact, there is a long history of studying the charge effects in compacted objects, especially in neutron stars and
 quark stars \cite{Bekenstein71,Dionysiou82,Lemos03,Ghezzi05,Lun07,Usov09}. One important conclusion is that a net charge distribution can support a higher
  maximum  mass of compact star
 significantly. While in white dwarf, there are also several authors concern the
charge effects, for instance in reference \cite{Olson76}, a two components
charge fluid model of white dwarf has been proposed. However, in
\cite{Olson76} they still assume that the net charge of white
dwarf is very tiny as people generally believed. Along this line,
here we propose a new mechanism: assuming existence of strongly
charged white dwarf to support the super-Chandrasekhar white
dwarf. Although this assumption has not yet confirmed by
observation, it still can serve as a useful toy model, especially in the theoretical discussion. In this new picture, we believe that
most of the white dwarf is nearly neutral and the Chandrasekhar mass limit still remains valid for most of the white dwarf, just as the astro observations indicated.
However, there may exist some white dwarfs, if enough net charge distributes in the star, as the force between the charged particles is repulsive and thus
equivalently can stiffen the equation of state of the white dwarf matters, then the maximum mass
of the charged white dwarf should  be significantly enlarged. Therefore, this in turn can provide a natural mechanism to
explain the peculiar type Ia supernova explosion as  mentioned in the beginning.

Contrast with the magnetic field effect, one of the most striking
feature of charged white dwarf is its dynamical stability. The
major difference between electric field and magnetic field is that
there exists an exactly spherical symmetry electrostatic field,
  while a  spherical symmetric magnetostatic field is not even existed. Thus unlike the magnetized white dwarf, the structure of charged white
 dwarf will not be deformed by the non-spherical symmetric field.

This paper is organized as follows: After a short introduction, we
present the basic structure formulation of the charged white dwarf and the corresponding numerical results in
Sec. II, where two subsections are included. The exact solution of the charged white dwarf is obtained in subsection A,
and the numerical results are shown in subsection B.
The stability issue of the charged white dwarf is examined in Sec. III, and the general relativistic effects on the charged white dwarf are shown in Sec. IV. Conclusions and outlooks are given in the last section.

\section{Hydrostatic Equilibrium Equation of Charged White Dwarf}

 For a static and spherical symmetric
charged white dwarf,  the outward force induced by the degenerate
electron gas and the repulsive electrostatic force caused by the
net charge are balanced by the inward gravitational force. In the
Newtonian framework, the hydrostatic equilibrium equation   can be
written as
\begin{equation}\label{dp}
\frac{dp}{dr}=-\frac{Gm\rho}{r^2}+\frac{qdq}{16\pi^2\varepsilon_0r^4dr},
\end{equation}
where $p$ and $\rho$ are the pressure and the mass density,
respectively; the net charges $q$ inside the radius $r$ can be
calculated by
\begin{equation}\label{}
q(r)=\int_0^r4\pi\rho_q r^2dr,
\end{equation}
 in which $\rho_q$ is the charge density; and the stellar
mass $m$ inside the radius $r$ is related with the  matter density
through
\begin{equation}\label{dm}
\frac{dm}{dr}=4\pi r^2\rho.
\end{equation}
 In addition, we use the capital letters $M$, $Q$ denote the total mass and total charge of a charged white dwarf.
In order to simplify the discussion, we assume the charge density
of white dwarf proportional to its matter density, ie
\begin{equation}\label{rhoq}
\rho_q=\alpha\times\frac{e}{m_{_p}}\times\rho,
\end{equation} where $\alpha$ is a dimensionless constant, $e$ and $m_{_p}$ are the charge and   mass of a proton, respectively.
Note that this kind of charge distribution is also adopted by
many other authors\cite{Lemos03,Usov09}. Moreover, since our electric field can be regarded as a particular case of anisotropic matter\cite{HB13a,HB13b}, therefore the following subsection A can be viewed as a direct application of the general formalism of polytropic Newtonian star with anisotropic pressure developed by Herrera and Barreto \cite{HB13a}.

\subsection{exact solution for polytropic equation of state}

Combining Eqs. (\ref{dp})-(\ref{dm}), through a careful and
complicated calculation, we obtain an exact solution by choosing
the equation of state with a form of polytropic type
$p=\kappa\rho^\gamma=\kappa\rho^{1+\frac1n}$, where $n$ is an
integer. By using the
 charge distribution of Eq. (\ref{rhoq}), the equilibrium equation (\ref{dp}) can be reduced to
\ba
\frac{dp}{dr}&=&-\frac{Gm\rho}{r^2}\left[1-\frac{\alpha^2e^2}{Gm_{p}^24\pi\varepsilon_0}\right]\nn\\
&=&-\frac{Gm\rho}{r^2}\left[1-\alpha^2\beta\right], \ea where
$\beta=\frac{e^2}{Gm_{p}^24\pi\varepsilon_0}=1.24\times 10^{36}$
is also a dimensionless constant. Combining this result with
Eq.(\ref{dm}), we get \ba
\frac{1}{r^2}\frac{d}{dr}\left(\frac{r^2dp}{\rho dr}\right)=-4\pi
G\rho\left[1-\alpha^2\beta\right].\label{dp1}
 \ea Now we introduce two new variables defined by
 \ba
\rho=\rho_c\theta^n, \quad\quad
\xi=\frac{r}{a},\label{newvariables} \ea
 where $\rho_c$ is the central density of the white dwarf and
$a$ is a constant defined by \ba
a=\left[\frac{(n+1)\kappa\rho_c^{\frac{1}{n}-1}}{4\pi
G\left(1-\alpha^2\beta\right)}\right]^{\frac12}. \ea By using
these new defined variables and the polytropic form of equation of
state, Eq.(\ref{dp}) is reduced to the standard form of Lane-Emden
equation\cite{Weinberg72}
 \ba
\frac{d}{d\xi}\left(\xi^2\frac{d\theta}{d\xi}\right)+\theta^n\xi^2=0\label{LE}.
\ea
 By employing the standard boundary conditions
 \ba
\theta(\xi=0)=1,\quad\quad \frac{d\theta}{d\xi}\Big|_{\xi=0}=0,
 \ea
Eq.(\ref{LE}) can be solved analytically. The radius of white
dwarf relates to a finite value of $\xi=\xi_1$ with $R=a\xi_1$,
where $\xi_1$ is the first zero point of $\theta$ function.
Meanwhile, the mass of white dwarf can be obtained by combining
Eqs. (\ref{dm}),(\ref{newvariables}),(\ref{LE}) \ba
M&=&\int_0^R4\pi r^2\rho(r)dr\nn\\
&=&4\pi a^3\rho_c\int_0^{\xi_1}\xi^2\theta^nd\xi\nn\\
&=&4\pi a^3\rho_c\xi_1^2\abs{\theta'(\xi_1)}. \ea
It is worth
noting that when the index $\gamma=\frac43$ (or equivalently
$n=3$), the mass of charged white dwarf becomes independent with
$\rho_c$,
and thus corresponds with the maximal mass of charged white dwarf. By comparing these results with the uncharged white dwarf, it is easy to see that the
difference is only reflected by the definition of $a$, i.e. we have
  \ba
R=\frac{R_{Ch}}{\left(1-\alpha^2\beta\right)^\frac12},\quad\quad M=\frac{M_{Ch}}{\left(1-\alpha^2\beta\right)^\frac32}\label{MR}
\ea
 where $R_{Ch}$ and $M_{Ch}$ are the Chandrasekhar radius and Chandrasekhar mass limit for uncharged white dwarf, respectively. Finally,
 it is worth noting that when $\alpha=0$ the Chandrasekhar mass limit of uncharged white dwarf will be recovered perfectly.

\subsection{numerical solution for general equation of state}

It is generally believed that the equation of state of white dwarfs can be simply described by the equation of state of the free electron gas
\begin{equation}\label{kF}
k_{_F}=\hbar(\frac{3\pi^2\rho}{m_p\mu})^{1/3},
\end{equation}

\begin{equation}\label{p2}
p=\frac{8\pi c}{3(2\pi\hbar)^3}\int_0^{k_F}\frac{k^2}{(k^2+m_e^2c^2)^{1/2}}k^2dk
\end{equation}
where $k$ represents the momentum of electrons, $k_F$ is the
maximum momentum determined by mass density, $\mu$ is the ratio of
nucleon numbers to electron numbers. It is easy to see that even
in this simple case, the equation of state still can not be
attributed to polytropic type, and thus become difficult to be
solved analytically. In view of these facts,  the equations
(\ref{dp})-(\ref{dm}) should be treated by numerical calculation.

\begin{figure}\label{MR0}
\centering
\includegraphics [width=0.7\textwidth]{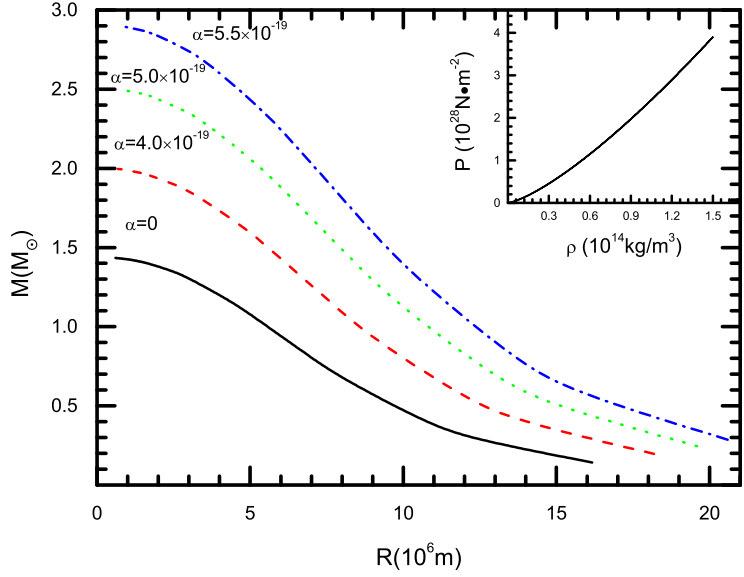}
\caption{\label{mr0} The mass-radius relations of the charged
white dwarf, where the different lines represent the different
value of $\alpha$  and the central density ranges from
$5\times10^8kg/m^3$ to $1.37\times10^{14}kg/m^3$. The inset shows
the equation of state of  free electron gas.}
\end{figure}

Combining the above equation of state with the Newtonian
equilibrium equations (\ref{dp})-(\ref{dm}) and the boundary
conditions: $m=0$ at $r=0$ and $p=0$ at $r=R$, we can obtain the
mass-radius relation for charged white dwarf numerically. the
numerical results for the charged and uncharged white dwarf are
presented in Fig. \ref{mr0}. From this figure it is shown that at
a fixed radius, a higher charge density corresponds with a higher
stellar mass.
When the $\alpha$ takes a range from $4.0\times10^{-19}$ to $5.0\times10^{-19}$, the corresponding maximum mass of charged white dwarf
varies from $2.0M_\odot$ to $2.5M_\odot$ continuously. Hence this picture can naturally explain the recent observational data.

\begin{figure}
\centering
\includegraphics [width=0.7\textwidth]{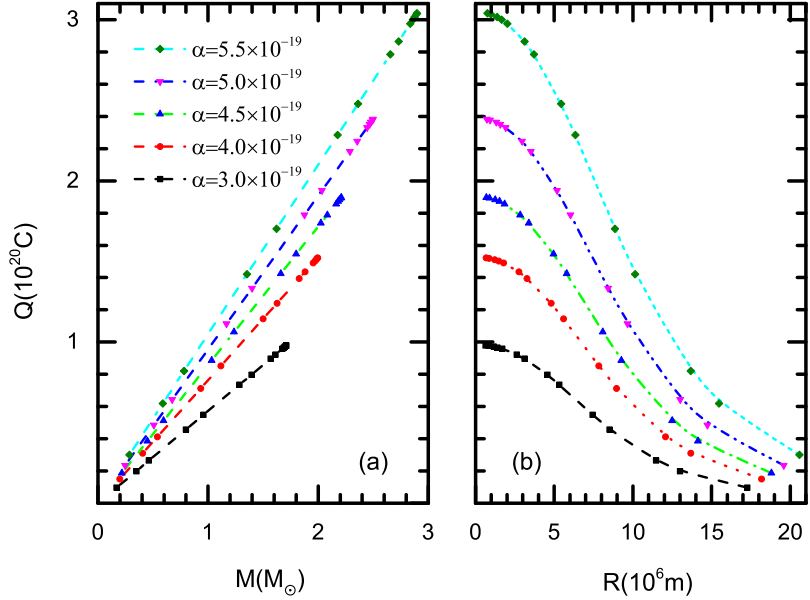}
\caption{\label{qm}  The Q-M relations  (a) and    the Q-R
relations (b) for different values of $\alpha$. }
\end{figure}

\begin{figure}
\centering
\includegraphics [width=0.7\textwidth]{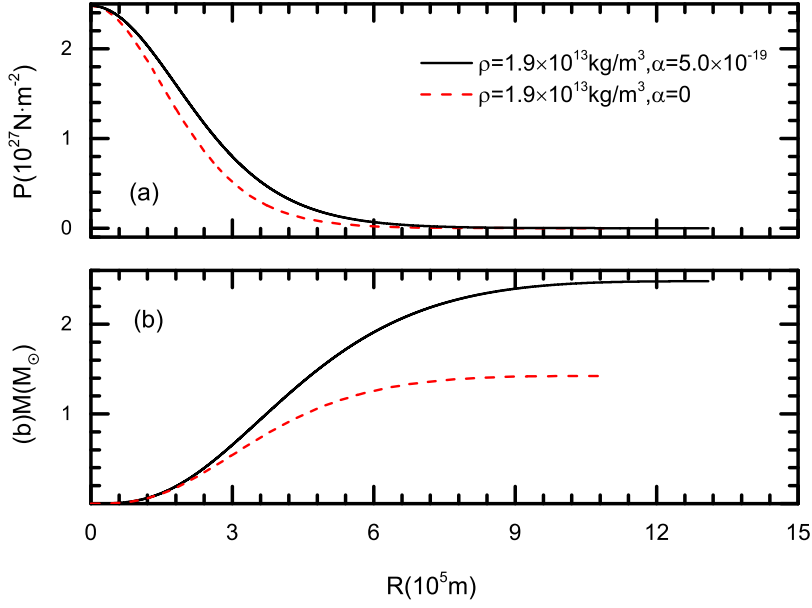}
\caption{\label{cc} The pressure distribution (a) and  mass
distribution (b) for the charged (solid lines) and uncharged
(dashed lines) white dwarf at a fixed central density
$\rho_c=1.9\times 10^{13}kg/m^3$.}
\end{figure}

\begin{figure}
\centering
\includegraphics [width=0.7\textwidth]{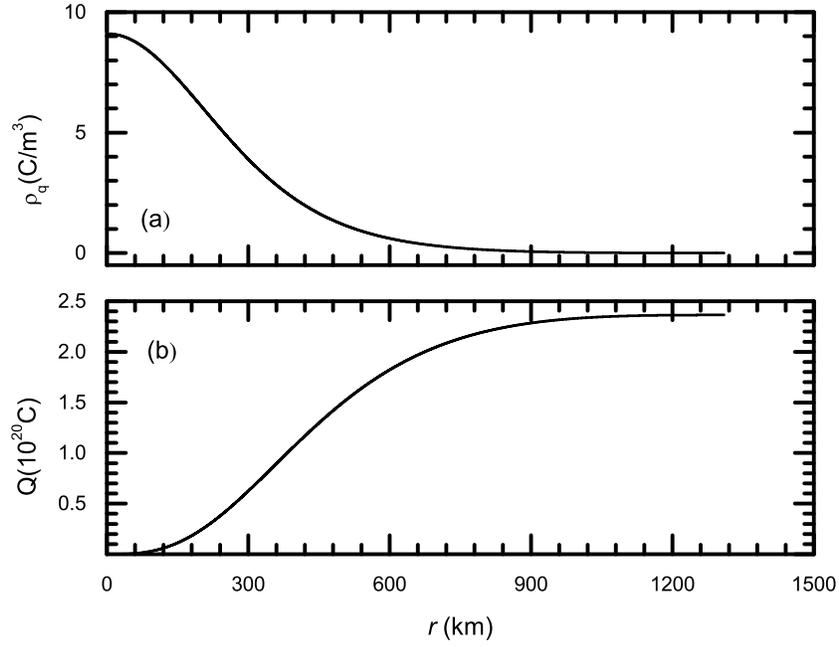}
\caption{\label{cd}The charge density  (a) and the total charge
(b) as   functions  of radius $r$, where $\rho_c=1.9\times
10^{13}kg/m^3$, and $\alpha=5.0\times 10^{-19}$.}
\end{figure}

\begin{figure}
\centering
\includegraphics [width=0.7\textwidth]{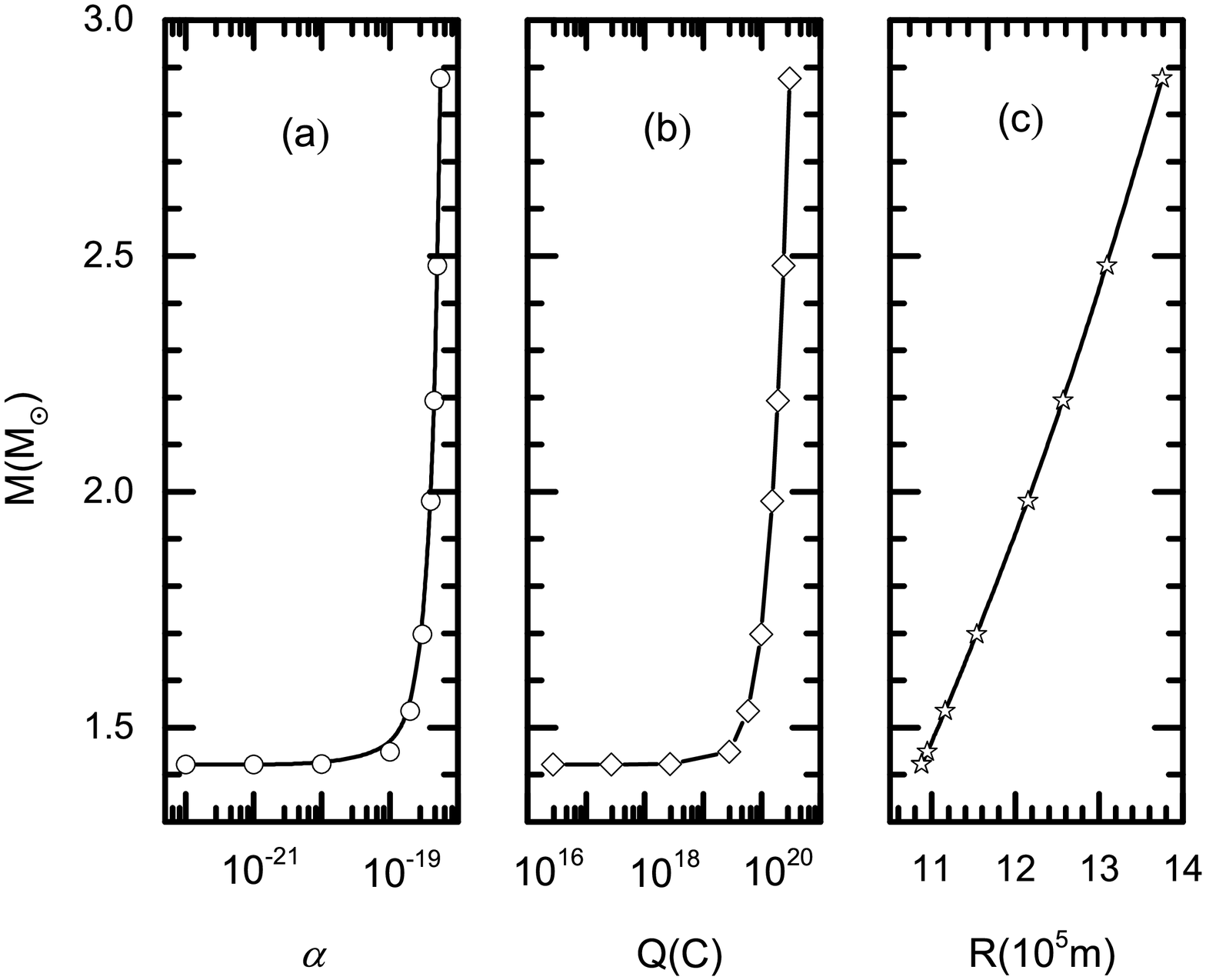}
\caption{\label{mq} The M-$\alpha$, M-Q   and M-R relations,
 where $\rho_c=1.9\times 10^{13}kg/m^3$.}
\end{figure}

In order to show the structures and properties of the charged
white dwarf intuitively, and also compare them to that of the
uncharged white dwarf, the numerical results are presented in the
Figs. (\ref{qm})-(\ref{mq}).  In Fig. \ref{qm}, the $Q$ (total charge)-$M$
(stellar mass) relations and the $Q$-$R$ (stellar radius)
relations are plotted.
 And in order to see the charge influence on a
single white dwarf more clearly, we plot the pressure and mass
distribution in Fig. \ref{cc} for $\rho_c=1.9\times 10^{13}kg/m^3$
and $\alpha=5.0\times 10^{-19}$, where the distributions of
corresponding uncharged white dwarf  are also plotted. Moreover,
 in Fig.
\ref{cd}  we   show the charge density and the total charge as
functions of radius $r$  at the same $\rho_c $ and $\alpha $ as
used in Fig. \ref{cc}. In the end,  the $M$-$\alpha$, $M$-$Q$ and
$M$-$R$ relations at  $\rho_c=1.9\times 10^{13}kg/m^3$ are shown in
Fig. \ref{mq}. It is interesting to note that when $\alpha$ is
smaller than $1\times10^{-20}$, the charge effects on the
mass-radius relation of white dwarf can be negligible. However,
when we further increase $\alpha$ to reach the order of
$10^{-19}$, the maximal mass of corresponding charged white dwarf
will become  sensitive to the value of $\alpha$, and will grow up
very rapidly as $\alpha$ increases.

\section{The Stability of The Charged White Dwarf}

Since we have already known that a super strong magnetic field in white
dwarf will cause the whole system dynamical
instability\cite{Zuo14,CFD13,Ruffini13},
 hence when one introduce a strong electric field in white dwarf, it is crucial to investigate the same stability issues.
 Based on the results in Sec. II, it is shown that the mass of
 charged white dwarf reaches its maximal value with a central density blow  than the neutralizing
 density\cite{CFD13}, thus
  our model can avoid the instability caused by neutralization.

On the other hand, in a seminal paper\cite{CF53}, Chandrasekhar
and Fermi generalize the Viral theorem   including the magnetic
field and it   in turn sets a constraint on the value of magnetic
field. Based on this famous result,  another kind of instability
of white dwarf
  with a super strong magnetic field is introduced by Ruffni et.al.\cite{Ruffini13}.
  Does this kind of instability also existed in white
  dwarf with super strong electric field? To answer this question, we use the similar strategy of reference \cite{CF53}, that is,  to generalize the Virial
  theorem
  including the electric field:

\begin{equation}\label{Th}
3\Pi+W_{_E}+W_{_G}=0.
\end{equation}
For a polytropic equation of state   $P=K\rho^{\gamma}$, we can
get $\Pi=(\gamma-1)U$, where $U$ is the internal energy. In order to proof this theorem, we start from the usual
assumptions of electrohydrodynamics \cite{Castellanos98,Zhakin12}.
The equations of motion governing an inviscid fluid can be written
in the form \cite{Zhakin12}

\ba \label{motion} \rho\frac{du_i}{dt}=-\frac{\partial}{\partial
x_i}(p+\frac{\varepsilon_r\varepsilon_0|E|^2}{2})
+\rho\frac{\partial \Omega}{\partial
x_i}+\varepsilon_r\varepsilon_0 \frac{\partial}{\partial
x_j}\left(E_i E_j\right) \ea
 where $\rho$ and $p$ denote the
density and pressure of the charged fuild respectively,
$\varepsilon_r$ is relative permittivity, $\Omega$ is the
gravitational potential, and $E$ is the intensity of the electric
field. Multiply Eq. (\ref{motion}) by $x_i$ and integrate over the
volume of the configuration, then the left-hand side of the
equation reads

\begin{equation}\label{}
    \begin{aligned}
\int \int \int \rho x_i \frac{du_i}{dt}dx_1 dx_2 dx_3 &= \int_0^M x_i\frac{d^2 x_i}{dt^2}dm\\
 &=\int_0^M \left[\frac{d}{dt}(x_i\frac{dx_i}{dt})-(\frac{dx_i}{dt})^2\right]dm\\
 &=\int_0^M\frac{d}{dt}(\frac{1}{2}\frac{dx_i^2}{dt})dm-\int_0^M(\frac{dx_i}{dt})^2dm\\
 &=\frac{1}{2}\frac{d^2}{dt^2}\int_0^Mr^2dm-\int_0^M|u|^2dm
    \end{aligned}
\end{equation}
where $dm=\rho dx_1dx_2dx_3$ and $M$ is the total mass of the
configuration. Letting

\begin{equation}\label{}
I=\int_0^Mr^2dm \qquad    \textrm{and}  \qquad
T=\frac{1}{2}\int|u|^2dm
\end{equation}
represent the moment of inertia and the kinetic energy of mass
motion, respectively, we can rewrite the Eq. (\ref{motion}) as

\begin{equation}\label{IT}
\frac{1}{2}\frac{d^2I}{dt^2}-2T=-\int\int\int
x_i\frac{\partial}{\partial
x_i}(p+\varepsilon_r\varepsilon_0\frac{|E|^2}{2})dx_1dx_2dx_3
+\varepsilon_r\varepsilon_0\int\int\int
x_i\frac{\partial}{\partial
x_j}\left(E_iE_j\right)dx_1dx_2dx_3+\int_0^Mx_i\frac{\partial
\Omega}{\partial x_i}dm.
\end{equation}
The last term of the right-hand side of this equation represents
the gravitational potential energy of the configuration:
 \ba
W_G&=&\int_0^Mx_i\frac{\partial \Omega}{\partial x_i}dm=-\int_0^M\frac{Gm(r)}{r}dm(r)\nn\\
&=&-\frac{3(\gamma-1)}{5\gamma-6}\frac{GM^2}{R}, \ea
 where the fact that the charged white dwarf configuration being spherical symmetric is employed. The remaining two terms of equation (\ref{IT})
 can also be reduced through
integration by parts

\begin{equation}\label{}
    \begin{aligned}
-\int\int\int x_i\frac{\partial}{\partial
x_i}(p+\varepsilon_r\varepsilon_0\frac{|E|^2}{2})dx_1dx_2dx_3
&=-\int\int\int\left[\frac{\partial}{\partial
x_i}\left(x_i(p+\varepsilon_r\varepsilon_0\frac{|E|^2}{2})\right)-
\left(\frac{\partial x_i}{\partial x_i}\right)(p+\varepsilon_r\varepsilon_0\frac{|E|^2}{2})\right]dx_1dx_2dx_3\\
&=-\int(p+\varepsilon_r\varepsilon_0\frac{|E|^2}{2})\textbf{r}\cdot
d\textbf{s}+3\int\int\int(p+\varepsilon_r\varepsilon_0\frac{|E|^2}{2})dx_1dx_2dx_3.
    \end{aligned}
\end{equation}
The surface integral over $d\textbf{s}$ vanishes as the pressure (including the electric pressure
$\varepsilon_r\varepsilon_0\frac{|E|^2}{2}$) must vanish on the
boundary of the configuration; and it is easy to see that the
volume integral over $p$ and
$\varepsilon_r\varepsilon_0\frac{|E|^2}{2}$ give the internal
energy ($U$) and the electric energy ($W_E$) of the configuration.
Thus we have

\begin{equation}\label{xxx}
-\int\int\int x_i \frac{\partial}{\partial
x_i}(p+\varepsilon_r\varepsilon_0\frac{|E|^2}{2})dx_1dx_2dx_3=3(\gamma-1)U+3W_E,
\end{equation}
where   the relation $p=(\gamma-1)\epsilon$ \cite{Weinberg72} is
used, and as $\epsilon=\rho-m_Nn$ is the density of internal energy, here $m_N=1.66\times10^{-27}kg$ being the rest mass per nucleon. Thus the
volume integral of $ \epsilon$ gives the total heat energy $U$. In
the same manner, the second volume integral in equation (\ref{IT})
gives

\begin{equation}\label{hh}
\varepsilon_r\varepsilon_0\int\int\int x_i\frac{\partial}{\partial
x_j}\left(E_iE_j\right)dx_1dx_2dx_3=-2W_E.
\end{equation}
Now, combining equations (\ref{IT})(\ref{xxx})(\ref{hh}), we have

\begin{equation}\label{}
\frac{1}{2}\frac{d^2I}{dt^2}=2T+3(\gamma-1)U+W_E+W_G
\end{equation}
This is just the generalized version of the Virial theorem
including electric field, it differs from the usual one only in
the appearance of $W_E+W_G$ in place of $W_G$.

By applying this generalized Virial theorem to a charged
non-rotating equilibrium star, then above equation reduce to
\begin{equation}\label{virial}
3(\gamma-1)U+W_E+W_G=0
\end{equation}
This is nothing but Eq.(\ref{Th}) we want to prove. With the help
of the expression of charge density described by Eq. (\ref{rhoq}),
 the electric energy $W_E$ appeared in the above equation can also be integrated
 exactly:
\ba\label{}
W_{_E}&=&\int_0^R\frac{1}{2}\varepsilon_0 (\frac{q(r)}{4\pi \varepsilon_0 r^2})^2 4\pi r^2 dr=\int_0^R\frac{q^2(r)}{8\pi\varepsilon_0 r^2}dr \nn\\
&=&\frac{\alpha^2e^2}{m_p^2}\int_0^R\frac{m^2(r)}{8\pi\varepsilon_0 r^2}dr \nn\\
&=&\frac{\alpha^2\beta}{2}\int_0^R\frac{Gm^2(r)}{ r^2}dr\nn\\
&=&\frac{\alpha^2\beta}{2}\frac{\gamma}{(5\gamma-6)}\frac{GM^2}{R}
\ea
 Here, for simplicity and without loss any generality, we assume relative permittivity $\varepsilon_r=1$. If $\varepsilon_r>1$,
 the electric field energy will be suppressed by a factor $\varepsilon_r$ and thus such a $\varepsilon_r$ will make the whole system more stable. Now we define
the total energy of the configuration by
\begin{equation}\label{Er}
W_{\textrm{total}}=U+W_{_E}+W_{G},
\end{equation}
as the gravitational potential energy of the configuration $W_G$
is negative, in order to form a bounded star configuration,
 the value of other positive energy such as electric energy and heat energy, should not be larger than the value of the negative gravitational
  potential energy. Hence similar to the results of reference \cite{CF53}, a necessary condition for the dynamical stability of a charged white dwarf
  equilibrium configuration should be $W_{total}<0$.
By eliminating $U$ between Eqs. (\ref{Th}) and (\ref{Er}), one can
obtain

\ba\label{}
W_{total}&=&-\frac{3\gamma-4}{3(\gamma-1)}(|W_{_G}|-W_{_E}).
 \ea
According to above equation, one can find that  if the electric
energy is larger than the gravitational potential energy,
 even for $\gamma>\frac43$( the condition of dynamical stability in the absence of electric field\cite{Weinberg72}),
 the whole system can also be dynamical instability.
 However, we can see that this is not the case, as this will require

 \ba
1<\frac{W_E}{|W_{_G}|}=\frac{\frac{\alpha^2\beta}{2}\gamma}{3(\gamma-1)},
\ea
 combining this condition with $\gamma>\frac43$ led to $\alpha^2\beta>\frac32$, however, this is inconsistence with Eq.(\ref{MR}) which states $\alpha^2\beta<1$.
Hence, in conclusion, for $\gamma>\frac43$, the charged white
dwarf is dynamically stable. In addition, it is easy to
 see that when $\alpha=0$ (i.e. uncharged white dwarf), our conclusion is nicely coincide with the well known result in reference \cite{Weinberg72}.

\section{General Relativity Effects of The Charged White Dwarf}

\begin{figure}
\centering
\includegraphics [width=0.7\textwidth]{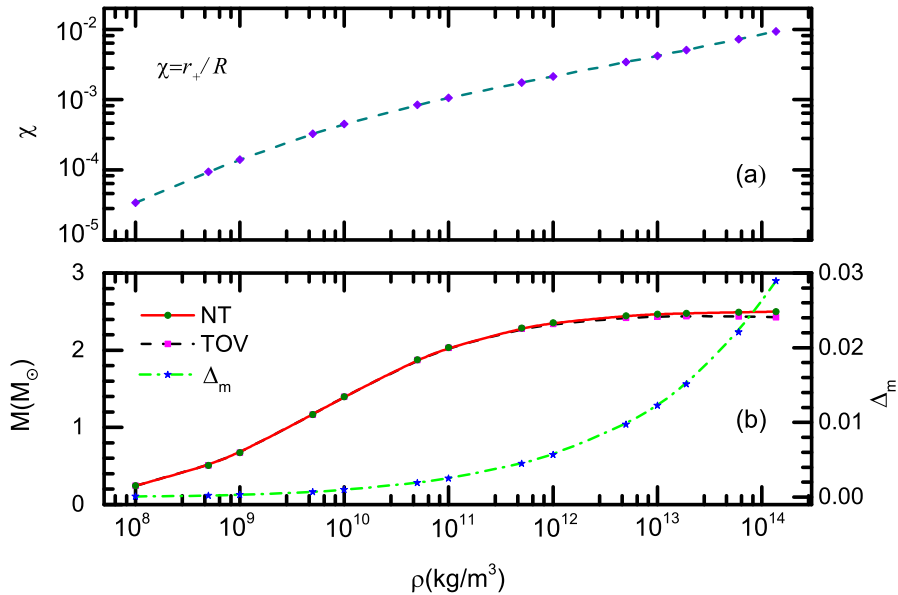}
\caption{\label{GR} The general relativistic effect in charged
white dwarfs, where the solid   line and the dash-dot   line in
(b) represent the stellar masses   under the framework of
Newtonian  gravity and the general relativity, respectively,
 while the green line in (b) represents the mass difference $\Delta_m$ between the two gravity theories, where $\alpha$=$5.0\times10^{-19}$. }
\end{figure}

\begin{table}
\caption{\label{tab1}$\alpha$, $R$, $M_{NT}/M_{\odot}$,
$M_{TOV}/M_{\odot}$, $\Delta_m$, $\chi$ of the charged white dwarfs
with $\rho_c=1.9\times10^{13}kg/m^3$.}
\begin{ruledtabular}
\begin{tabular}{cccccc}
$\alpha$ & $R(km)$ & $M_{NT}/M_\odot$ & $M_{TOV}/M_\odot$ & $\Delta_m$ & $\chi$\\
\hline
$8.6\times10^{-19}$ & $3754.8$ & $58.4744$ & $52.7244$ & $0.09833$ & $0.02959$\\
\hline
$8.5\times10^{-19}$ & $3355.3$ & $41.7256$ & $38.3719$ & $0.08038$ & $0.02426$\\
\hline
$8.3\times10^{-19}$ & $2839.9$ & $25.3008$ & $23.7974$ & $0.05942$ & $0.01816$\\
\hline
$8.0\times10^{-19}$ & $2389.3$ & $15.0675$ & $14.4134$ & $0.04341$ & $0.01353$\\
\hline
$7.5\times10^{-19}$ & $1975.3$ & $8.5145$ & $8.2523$ & $0.03079$ & $0.00985$\\
\hline
$7.0\times10^{-19}$ & $1735.1$ & $5.7705$ & $5.6293$ & $0.02447$ & $0.00798$\\
\hline
$6.5\times10^{-19}$ & $1575.6$ & $4.3211$ & $4.2317$ & $0.02069$ & $0.00683$\\
\hline
$6.0\times10^{-19}$ & $1461.4$ & $3.4480$ & $3.3853$ & $0.01818$ & $0.00607$\\
\hline
$5.7\times10^{-19}$ & $1407.2$ & $3.0786$ & $3.0261$ & $0.01705$ & $0.00572$\\
\hline
$5.5\times10^{-19}$ & $1375.6$ & $2.8760$ & $2.8288$ & $0.01641$ & $0.00552$\\
\hline
$5.0\times10^{-19}$ & $1309.2$ & $2.4794$ & $2.4419$ & $0.01512$ & $0.00511$\\
\hline
$4.5\times10^{-19}$ & $1256.8$ & $2.1932$ & $2.1621$ & $0.01418$ & $0.00480$\\
\hline
$4.0\times10^{-19}$ & $1214.8$ & $1.9808$ & $1.9542$ & $0.01343$ & $0.00456$\\
\hline
$3.5\times10^{-19}$ & $1181.1$ & $1.8202$ & $1.7970$ & $0.01275$ & $0.00436$\\
\hline
$3.0\times10^{-19}$ & $1154.0$ & $1.6980$ & $1.6771$ & $0.01231$ & $0.00421$\\
\hline
$2.0\times10^{-19}$ & $1115.8$ & $1.5348$ & $1.5168$ & $0.01173$ & $0.00400$\\
\hline
$1.0\times10^{-19}$ & $1094.6$ & $1.4490$ & $1.4325$ & $0.01139$ & $0.00389$\\
\hline
$1.0\times10^{-20}$ & $1087.9$ & $1.4224$ & $1.4064$ & $0.01125$ & $0.00385$\\
\end{tabular}
\end{ruledtabular}
\end{table}

\begin{figure}
\centering
\includegraphics [width=0.7\textwidth]{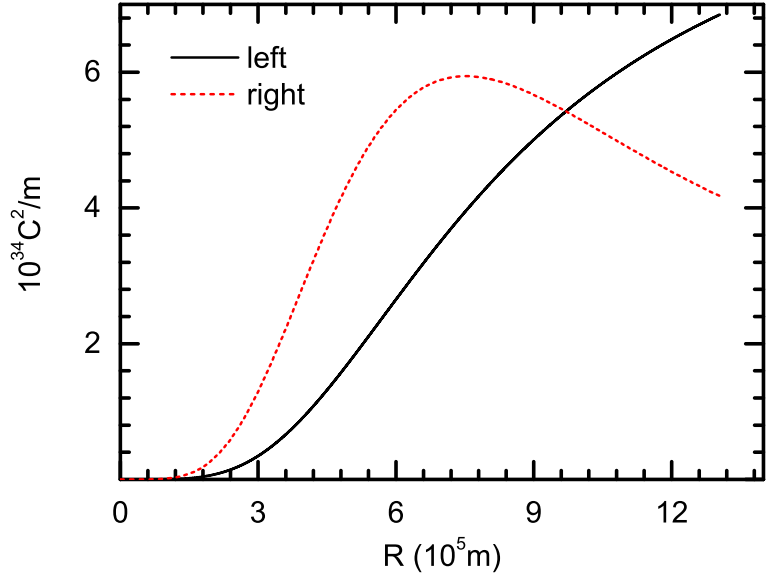}
\caption{\label{lr} The values of the left hand and right hand of Eq.(\ref{qsquare}) as a function of $r$.}
\end{figure}

\begin{figure}
\centering
\includegraphics [width=0.7\textwidth]{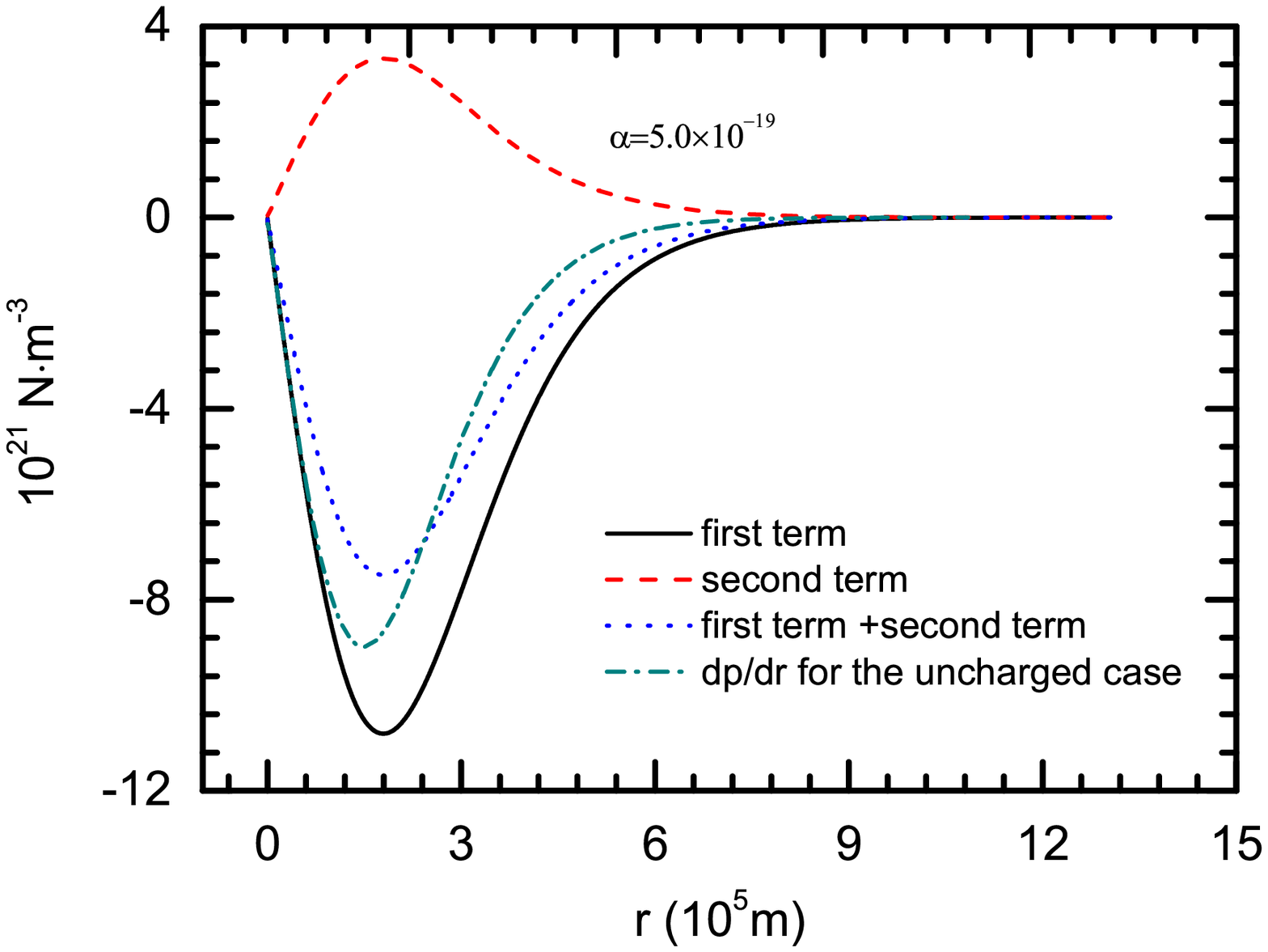}
\caption{\label{fig.dpdr} The right hand side terms of Eq. (\ref{dpdr}) with central density $\rho_c=1.9\times10^{13}kg/m^3$ and $\alpha=5.0\times10^{-19}$. The solid and dashed lines represent the first and second terms of right hand side of Eq. (\ref{dpdr}) respectively, the dotted line represents the overall effect of right hand side of Eq. (\ref{dpdr}), while the dot-dashed line represents $\frac{dp}{dr}$ of uncharged white dwarf ($\alpha=0$) with the same central density.}
\end{figure}

In the last section, the mass of charged white dwarf can reach up
to $3M_\odot$, which is much larger than the Chandrasekhar mass
limit. Thus it is natural to ask the validity of using Newtonian
hydrostatic equilibrium equations. Does the general relativity
effect become relevant and spoil all the results we obtained in
the above section? In this section, we will try to answer these
questions.

In general relativistic case, the spacetime outside of a charged
white dwarf can be described by the Reissner-Nordstrom
 solution \cite{Weinberg72}
  \ba
ds^2=-\left(1-\frac{2GM}{c^2r}+\frac{GQ^2(r)}{4\pi\varepsilon_0c^4r^2}\right)dt^2+\left(1-\frac{2GM}{c^2r}+\frac{GQ^2(r)}{4\pi\varepsilon_0c^4r^2}\right)^{-1}dr^2
+r^2\left(d\theta^2+\sin^2\theta d\varphi^2\right);
 \ea
Usually, one use the quantity $\chi=\frac{r_+}{R}$ to indicate the significance of general relativity effects, here $R$ denotes radius of charged white dwarf,
and $r_+=\frac{GM}{c^2}+\sqrt{\frac{G^2M^2}{c^4}-\frac{GQ^2}{4\pi\varepsilon_0c^4}}$ being the out horizon of Reissner-Nordstrom solution.
When $\alpha=5.0\times10^{-19}$, from the Fig. 7, we yield  $\chi_{max}=0.00936$, which is very small, and hence in turn implies the general
relativity effects in this case can be negligible. However, in order to justify this roughly estimates and concrete our result, we perform a detailed
 analysis on general relativity effects of charged white dwarf and to see how much it deviates from Newtonian dynamics.
The
interior line element of the static and spherical symmetric
charged white dwarf reads
 \ba
 ds^2=-e^\nu dt^2+e^\lambda dr^2+r^2\left(d\theta^2+\sin^2\theta
 d\varphi^2\right),
 \ea
where $\nu$ and $\lambda$ are only the function of radius $r$. And
the classical hydrostatic equilibrium equations
(\ref{dp})-(\ref{dm}) should be replaced by \cite{Bekenstein71}

\begin{equation}\label{dpdr}
\frac{dp}{dr}=-\frac{p+\rho c^2}{r(r-\frac{2Gm}{c^2}+\frac{Gq^2}{4\pi\varepsilon_0c^4r})}\left[\frac{4\pi G}{c^4}pr^3+\frac{Gm}{c^2}
-\frac{Gq^2}{4\pi\varepsilon_0c^4r}\right]+\frac{qdq}{16\pi^2\varepsilon_0r^4dr},
\end{equation}
\begin{equation}\label{}
\frac{dm}{dr}=4\pi r^2\rho+\frac{qdq}{4\pi\varepsilon_0c^2rdr},
\end{equation}
and
\begin{equation}\label{} q=\int_0^r4\pi\rho_q
e^{\lambda/2}r^2dr,
\end{equation}
 where
 \ba
e^{-\lambda}=1-\frac{2Gm(r)}{c^2r}+\frac{Gq^2(r)}{4\pi\varepsilon_0c^4r^2}.\label{elambda}
\ea
 By solving those equations numerically, we can compare our
results with the Newtonian case. More specifically, we define a
quantity $\Delta_m=\frac{M_{NT}-M_{TOV}}{M_{NT}}$, where $M_{NT}$
and $M_{TOV}$ represent the mass of white dwarf in the framework
of
 Newtonian gravity and general relativity, respectively.

The numerical results are presented in Fig. \ref{GR}
and Tab. \ref{tab1}. It is shown  that the general relativity effect is indeed negligible
when the mass of charged white dwarf smaller than $3M_\odot$. If we further increase the value of $\alpha$, the mass of corresponding charged white dwarf will
became more higher, and the general relativistic effects will became relevant.

It is worthwhile to note that \cite{Bekenstein71,Prisco07} if the condition\ba
\int^r_0\frac{q^2}{s^2}ds>\frac{q^2}{r}\label{qsquare}
\ea is satisfied, the so called ``active gravitational mass" \cite{Prisco07} will be increased, and hence contribute to the gravitational binding of the sphere. Then people natural wondering whether this interesting effect will decrease the effective mass of charged white dwarf? To illustrate this problem more clearly, we take a particular charged white dwarf with central matter density $\rho_c=1.9\times10^{13}kg/m^3$ and $\alpha=5.0\times10^{-19}$ as an example. From Fig. \ref{lr} we can see that when $r>9.70\times10^5$$m$ the left hand side of Eq.(\ref{qsquare}) is larger than the right hand side, and thus implies the first term of right hand side of Eq.(\ref{dpdr}) becomes more negative. However, since the second term of the right hand side of Eq. (\ref{dpdr}) (i.e. $\frac{qdq}{16\pi^2\varepsilon_0r^4dr}$) is always positive. Therefore the gradient of pressure of charged white dwarf is less negative than the uncharged white dwarf when $r<2.42\times10^5$$m$ as shown in Fig. \ref{fig.dpdr}. And this in turn ensure the mass of charged white dwarf with $\alpha=5.0\times10^{-19}$ is uplifted to $2.44M_\odot$.

\section{Conclusions}
In this paper, in order to explain  some peculiar Type Ia
supernova explosion, we propose a new mechanism based on the
existence of charged white dwarf. Our calculations show that the
mass of charged white dwarf can be uplifted significantly, and
hence can naturally explain the observed peculiar Type Ia
supernova explosion. Particularly, by employing a representative choice of charge distribution, we obtain an analytic
solution for the stellar structure, as shown in Eq.  ({\ref{MR}}).
Then we continue to investigate the
stability issue, the detailed calculations indicate that the charged white dwarf configuration is stable when the polytropic index $\gamma>\frac43$.
Since our treatment is based on Newtonian framework, to ensure the reliability of our results and also to explore the valid region of Newtonian framework,
we further investigate the general relativistic effects. Our investigation shows that the general relativistic effects can be negligible when
the mass of charged white dwarf is smaller than $3M_\odot$, and thus under this stellar mass the Newtonian framework is reliable.

It is worth noting that there are still many aspects of this
theory deserving further investigation. For example, charge
distribution other than Eq.(\ref{rhoq}) is of course allow in
principle, can we find another
charge distribution to reduce the total charge significantly?
Furthermore, since we require the existence of very strong
electric field in the charged white dwarf, is there any intrinsic observable
effects induced by those strong electric field? We hope to back those issues in the near future.

\begin{acknowledgements}
This work is supported by NSFC ( Nos.10947023, 11275073 and
11305063) and the Fundamental Research Funds for the Central
University of China under Grants No.2013ZG0036 and No.2013ZM107. This
project is sponsored by SRF for ROCS and SEM and has made use of NASA's
Astrophysics Data System.

\end{acknowledgements}


\end{document}